# Cross Layer QoS Support Architecture with Integrated CAC and Scheduling Algorithms for WiMAX BWA Networks

Prasun Chowdhury, Iti Saha Misra, Salil K Sanyal

Department of Electronics and Telecommunication Engineering
Jadavpur University, Kolkata-700032, India

*Abstract*— **In this paper, a new technique for cross layer design, based on present Eb/N0 (bit energy per noise density) ratio of the connections and target values of the Quality of Service (QoS) information parameters from MAC layer, is proposed to dynamically select the Modulation and Coding Scheme (MCS) at the PHY layer for WiMAX Broadband Wireless Access (BWA) networks. The QoS information parameter includes New Connection Blocking Probability (NCBP), Hand off Connection Dropping Probability (HCDP) and Connection Outage Probability (COP). In addition, a Signal to Interference plus Noise Ratio (SINR) based Call Admission Control (CAC) algorithm and Queue based Scheduling algorithm are integrated for the cross layer design. An analytical model using the Continuous Time Markov Chain (CTMC) is developed for performance evaluation of the algorithms under various MCS. The effect of Eb/No is observed for QoS information parameters in order to determine its optimum range. Simulation results show that the integrated CAC and packet Scheduling model maximizes the bandwidth utilization and fair allocation of the system resources for all types of MCS and guarantees the QoS to the connections.**

## I. INTRODUCTION

Successful delivery of real-time multimedia traffic over the IP based networks, like Internet is a challenging task because of the strict requirement of Quality-of-Service (QoS). Traditional IP based best effort service will be unable to fully meet all the stringent requirements. The time-varying nature of channels along with the resource constrained devices associated with wireless networks make the problem even more complicated. In wireless networks in contrast to wired networks, the channel impairments vary rapidly due to the fluctuation of channel conditions even when the transmitter and receiver are stationary. It is, therefore, desirable to adjust the transmitted power, type of modulation and data rate to match the channel condition dynamically in the best possible manner in order to support the highest possible channel capacity even under the worst case condition of the channel. To make it feasible, the cross layer design for wireless multimedia communication networks has gained significant popularity in the research community [1]. In cross layer design, the challenges from the physical wireless medium and the associated QoS demands for the relevant applications are taken into consideration so that the rate, power, and coding at the physical layer can adapt dynamically to meet the stringent requirements of the applications under the current channel and network conditions.

The ever-increasing wireless traffic is a conglomeration of various real-time traffic such as voice, multimedia teleconferencing, games and data traffic such as WWW browsing, messaging and file transfers etc. All these applications require widely varying and diverse QoS guarantees for different types of traffic. Next generation wireless networks such as WiMAX (Worldwide Interoperability for Microwave Access) or IEEE 802.16 [2] have different types of services with varied QoS requirements, but the IEEE 802.16 has not yet defined the standards for QoS guarantees. Wireless channels suffer from bandwidth limitation, fluctuations of the available bandwidth, packet loss, delay and jitter. Real-time media such as video and audio is highly delay sensitive but Non-real time media such as web data is comparatively less delay sensitive but both types require reliable data transmission. Some advanced techniques such as Orthogonal Frequency Division Multiple Access (OFDMA) [3], Adaptive Modulation and Coding (AMC) [4-6] and wide variety of protocols and standards [7] are used to combat the challenges for stringent QoS requirement in wireless networks. Also, the mobile devices are power constrained. Maintaining good channel quality in one side and minimizing average power consumption for processing and communication on the other are two very conflicting requirements. Receivers in multimedia delivery systems are quite different in terms of latency requirements, visual quality requirements, processing capabilities, power limitations, and bandwidth constraints. In view of the above constraints, a strict modularity and protocol layer independence of the traditional TCP/IP or OSI stack will lead to a sub-optimal performance of applications over IP based wireless networks [8]. For optimization, we require suitable protocol architectures that would modify the reference-layered stack by allowing direct communication between the protocols at non-adjacent layers or sharing of the state variables across different layers to achieve better performance. The objective of a cross layer design is to actively exploit this possible dependence between the protocol layers to achieve performance gains. Basically, a cross layer design involves feedback received from other layers to optimize the relevant parameters in the current layer to achieve the optimum performance. Although the cross layer design is an evolving area of research, considerable amount of work has already been done in this area [1].

In this paper, information from the MAC layer is used to optimize the parameter at PHY layer. Among several alternatives, the IEEE 802.16 PHY layer standard realizes the usage of Orthogonal Frequency Division Multiplexing (OFDM) in order to mitigate the adverse effects of frequency selective multi-path fading and to efficiently contrast Inter-Symbol and Inter Carrier Interferences (ISI and ICI) [9]. In addition, PHY layer also supports variable channel bandwidth, different FFT sizes, multiple cyclic prefix times and different modulation schemes such as QPSK, 16QAM and 64QAM with convolutional encoding at various coding rates [10]. On the other hand, Call Admission Control (CAC) mechanism [11-19] and Scheduling mechanism [17-23] are the important wings of WiMAX QoS framework at MAC layer. A CAC algorithm at base station (BS) will admit a subscriber station (SS) into the network if the BS ensures the minimum QoS requirement of the SS without degrading the existing QoS of other SSs in the network.

Providing CAC in IEEE 802.16 BWA networks guarantees the necessary QoS of different types of connections and at the same time decreases the Blocking Probability (BP) [13], Dropping Probability (DP) [14] and Outage Probability (OP) [24] for all types of connections. These QoS parameters can be used as feedback information to the PHY layer in order to select an appropriate Modulation and Coding Scheme (MCS) for better network performance. Again, scheduling algorithm at SS will distribute the bandwidth of the selected MCS among the real-time and the non-real-time traffic, based on the type of users in the network and their QoS requirements. Scheduling in the uplink direction is tedious because of required bandwidth request, bandwidth grant and transmission of data of each flow which affect the real-time application due to large Round Trip Delay (RTD) [25, 26].

The scheduling algorithm involves in fair allocation of the bandwidth among the users, determining their transmission order and enhancement of bandwidth utilization. Fairness refers to the equal allocation of network resources among the various users operating in both good and bad channel states. In this paper, fairness has been quantified using Jain's Fairness Index [27].

### A. Background literature

A significant number of proposals on cross layer designs have been found in recent literature. Many of them refer to feedback mechanisms between the PHY and MAC layers. In [28] authors have introduced a cross layer approach for calculating the QoS indicators (blocking rates, download time and bit error rates) taking into account the PHY layer conditions (modulation and coding, propagation and MIMO), the MAC layer radio resource management algorithms and the higher layer traffic characteristics. Authors have also considered different admission control schemes and studied the impact of adaptive modulation and coding on the performance of elastic traffic. But no packet scheduling algorithms for fair allocation of bandwidth among the traffic sources have been considered. Also optimum range of Signal to Noise Ratio (SNR) for adaptive MCS in the networks has not been determined.

A framework of memory less scheduling policies based on channel quality states, buffer occupancy states and retransmission number states has been provided in [29]. The scheduling policies combine AMC (Adaptive Modulation and Coding) and ARQ (Automatic Repeat Request) in a cross layer fashion that produces a good throughput performance with reduced average delay. However, the authors have not considered any CAC algorithm for connection management in the networks.

Similar to [29], [30] proposes a scheduling algorithm only at the MAC layer for multiple connections with diverse QoS requirements, where each connection employs AMC scheme at the PHY layer over wireless fading channels. A priority function is defined for each connection admitted in the system, which is updated dynamically depending on the wireless channel quality, QoS satisfaction, and services across all layers. The connection with highest priority is scheduled at each time. At the MAC layer, each connection belongs to a single service class and is associated with a set of QoS parameters that quantify its characteristics.

A cross layer optimization mechanism for multimedia traffic over IEEE 802.16 BWA networks has been proposed in [31, 32].The main functionality of the cross layer optimization mechanism resides at the BS part. The authors have used a decision algorithm at the BS part that relies on the values of two major QoS parameters, i.e., the packet loss rate and the mean delay. These QoS parameters activate proper adjustment of the modulation and/or the media coding rate, aiming at improving QoS and system throughput. However, this paper has not considered proper CAC and scheduling mechanisms for the delay sensitive real-time traffic flow at MAC layer. The authors have also not evaluated the optimum range of SNR in order to uniformly control modulation and data encoding rates.

A cross layer scheduler that employs AMC scheme at the PHY layer, according to the SNR on wireless fading channels has been described in [33]. The authors have defined cost function for each kind of multimedia connections based on its service status, throughput or deadline in MAC layer to achieve an optimum tradeoff between the throughput and fairness. Though the authors have included a suitable scheduling algorithm at the MAC layer but the authors have not considered any CAC algorithm.

[34] proposes a cross layer optimization architecture for WiMAX system. It consists of a Cross layer Optimizer (CLO), which acts as an interface between MAC and PHY layers. The CLO gathers and optimizes the parameters from both layers to achieve optimum performance gain. The CLO gets the channel condition information from the PHY layer and bandwidth requests and queue length information from the MAC layer so that it can switch to different burst profile for different modulation and coding schemes. In this case also, the authors have not specified proper CAC and scheduling mechanism at the MAC layer. Moreover, no optimum range of SNR has been taken into account in CLO while switching to different MCS.

Another cross layer approach as provided in [3], analyses several QoS parameters that include the bit rate and the bit error rate (BER) in the PHY layer, and packet average throughput/delay and packet maximum delay in the link layer. Authors show that dynamic OFDMA has a stronger potential to support multimedia transmission than dynamic OFDM-TDMA. Authors have taken care of proper scheduling, throughput and delay guarantee of traffic flows ignoring the CAC mechanism at MAC layer.

[4] describes a polling based uplink scheduling schemes for TCP based applications in a multipoint to-point fixed broadband IEEE 802.16 BWA network. This scheme adapts the transmission rates between the SSs and the BS dynamically using adaptive modulation technique. With adaptive modulation, the uplink scheduling algorithm helps to achieve higher data transmission rate, fairness in slot assignment and in the amount of data transmission.

### B. Contribution

The novelty of our work focuses on the dynamic selection of MCS based on Eb/N0 ratio of the concerned connections and relevant feedback information obtained from the MAC layer in terms of New Connection Blocking Probability (NCBP), Hand off Connection Dropping Probability (HCDP) and Connection Outage Probability (COP) to employ the most appropriate MCS in the network. The target values of the feedback information changes the selection of modulation and coding rate dynamically based on optimum range of Eb/N0 ratio of the connections to enable a flexible use of the network resources. The effect of Eb/N0 ratio has been observed on NCBP, HCDP and COP for all the relevant connection types under various MCS by means of exhaustive simulation. The simulation results help to determine the optimum range of Eb/N0 ratio for all the connections with appropriate MCS. It has been observed that the integrated CAC and Scheduling algorithm maximizes the utilization and fair allocation of the system resources for all types of MCS and provide better QoS support with reduced NCBP, HCDP and COP in the IEEE 802.16 BWA networks.

Since there is also a possibility that the wireless channel deteriorates or becomes unavailable at certain instant of time, the CAC and scheduling algorithms at the MAC layer, which has been left undefined in the standard, can get proper knowledge of the channel condition, in terms of SINR of the connections. Thereby, only the connection with a good channel condition is admitted and scheduled for transmission to achieve more gain in bandwidth utilization and fairness index.

An analytical model based on Continuous Time Markov Chain (CTMC) [35] is developed for the more realistic analysis of QoS parameters for the performance evaluation of IEEE 802.16 Broadband Wireless Access (BWA) Networks. As the Markov Chain basically involves state transition analysis, the probabilities for getting admission without degradation of QoS for both real as well as non-real time services is accurately analysed at the state level.

The remainder of the paper is organized as follows: Section II discusses System models. In addition to it, cross layer architecture and a joint algorithm for SINR based CAC and Queue based scheduling are proposed. Detailed description of the analytical model of proposed SINR based CAC algorithm is given in Section III. Section IV sets the parameter for performance evaluation. Section V shows QoS performance results numerically for different MCS. Finally, Section VI concludes the paper.

### II. SYSTEM MODEL AND CROSS LAYER ARCHITECTURE

To accommodate a wide variety of applications, WiMAX defines five scheduling services that should be supported by the BS MAC scheduler for data transport. Four service types are defined in IEEE 802.16d-2004 (Fixed) standard [12], which includes UGS (Unsolicited Grant Service), rtPS (Real-time Polling Service), nrtPS (Non Real-time Polling Service), and BE (Best Effort). In addition, one more service i.e. ertPS (Extended Real-time Polling Service) is defined in IEEE 802.16e-2005 (Mobile) standard [14]. The UGS is designed to support real-time service flow that generates fixed-size data periodically, such as T1/E1 and VoIP without silence suppression. On the other hand, the rtPS supports the same with variable data size, such as video streaming services. Similarly, the nrtPS deals with FTP. The BE and ertPS perform tasks related to e-mail and VoIP respectively. The guaranteed delay aspect is taken utmost care in video streaming and VoIP. In the mobile WiMAX environment, the handover procedure begins as soon as the mobile SS moves into the service range of another BS. The relevant QoS parameters specified in the standard are Maximum Sustained Traffic Rate (MSTR), Minimum Reserved Traffic Rate (MRTR), Maximum Latency (ML), Tolerated Jitter (TJ) and Request/Transmission Policy. The Service Flow Identifier (SFID), Connection Identifier (CID) and traffic priority are mandatory for all QoS classes [10].

### A. PHY Layer Model

IEEE 802.16 specifies multiple PHY specifications including Single Carrier (SC), SCa, OFDM and OFDMA. Among the several alternatives, we consider only OFDMA in the PHY802.16 layer for the cross layer design because of its limited interference in the network [9]. OFDMA is similar to OFDM using multiple sub-carriers to transmit data. However, while OFDM uses all available sub-carriers in each transmission, different sub-carriers could be arranged to different subscribers in downlink and each transmission could use the available sub-carriers in uplink in OFDMA.

We consider 802.16 PHY layer to be supported with variable channel bandwidth, different FFT sizes, multiple cyclic prefix time and different modulation schemes such as QPSK, 16QAM and 64QAM with convolutional encoding at various coding rates. The raw data rates of the OFDMA are functions of several parameters such as channel bandwidths, FFT size, sampling factor, cyclic prefix time, modulation scheme, encoding scheme and coding rate. It can be up to 70 Mbps by using high-grade modulation scheme with other suitable parameters [10, 36]. For proper design of PHY layer with MCS, we first observe the effect of all the said parameters in terms of raw data rate and spectrum efficiency.

In [36], a method of calculating raw data rate and OFDM symbol duration for different MCS is given. In this paper, we also derive a relevant term, spectrum efficiency (the alternative term bandwidth efficiency is also frequently used), being defined [37] as the transmitted bit per second per Hertz (b/s/Hz). This normalized quantity is a valuable system parameter. For instance, if data are transmitted at a rate of 1 Mbps in a 0.6 MHz wide baseband system, the spectral efficiency is 1 Mb/s/0.6 MHz, or 1.67 b/s/Hz. In this paper, spectral efficiency is evaluated as the ratio of raw data rate to the channel bandwidth as determined by the particular MCS in IEEE 802.16 BWA network. The calculation for the above parameters is given below. Let, 'R' be the raw data rate;

$$R = N * b * c / T \qquad (1)$$

where:

N = Number of used sub-carriers

b = Number of bits per modulation symbol

c = Coding rate

T = OFDM symbol duration

The value of 'N' is a function of the FFT size. Table I lists the values of 'N' for different FFT sizes.

TABLE I.     NUMBER OF USED SUB-CARRIERS AS A FUNCTION OF FFT SIZE [36]

| FFT Size | Number of Used Sub-carriers (N) |
|----------|--------------------------------|
| 2048 | 1440 |
| 1024 | 720 |
| 512 | 360 |
| 128 | 72 |

The OFDM symbol duration is obtained by using the following formulae [36]:

$$Fs = Sampling\ factor * Channel\ Bandwidth \qquad (2)$$
$$Tb = FFT\_size / Fs \qquad (3)$$
$$Tg = G * Tb \qquad (4)$$
$$T = Tb + Tg \qquad (5)$$

where:

Fs = Sampling frequency

Tb = Useful symbol time

Tg = Cyclic prefix time

G = Cyclic prefix

T = OFDM symbol duration

Sampling-Factor is set to 8/7, 'G' is varied as 1/32, 1/16, 1/8 and 1/4 and coding rates used are 1/2, 2/3, and 3/4. These are the standard values specified in WiMAX [36].

Based on the given values of Sampling-Factor, Channel bandwidth and 'G', the OFDM symbol duration 'T' is computed by using equations (2) to (5). Raw data rates can be computed using equation (1). We compute raw data rate and spectrum efficiency for different Cyclic Prefix, modulation schemes and coding rates, using FFT_size = 2048 and channel bandwidth= 20 MHz and the computed results are shown in Table II. From Table II, it is observed that raw data rate and spectrum efficiency are maximum when cyclic prefix G =1/32 (i.e. Minimum value of G) which is obvious because this value of G causes lower time spread of the signal which results in the larger data rate and hence provides maximum spectrum efficiency.

### B. Medium Access Control (MAC) Layer Model

We consider SINR based CAC algorithm for BS and Queue based Scheduling algorithm for SS in the WiMAX MAC layer for the design of cross layer model. The terminologies used in this section are given in Table III. We make the following assumptions in our proposed algorithm.

*1)* *ertPS connection requests are considered to be same as rtPS connections, because both connections have the same QoS parameters and differ only by the way of Request/Transmission policy.*

*2)* *BE connections are not considered in our CAC scheme, because they are designed to support best effort flows which do not need any QoS guarantees.*

*3)* *The connections of the similar service types have same QoS parameter values.*

TABLE II.     LIST OF DATA RATES AND SPECTRUM EFFICIENCY FOR DIFFERENT MCS AND CYCLIC PREFIX

| Modulation Type | Coding Rate | Cyclic Prefix=1/32 | | Cyclic Prefix=1/16 | | Cyclic Prefix=1/8 | | Cyclic Prefix=1/4 | |
|---|---|---|---|---|---|---|---|---|---|
| | | Raw data rate (Mbps) | Spectrum Efficiency | Raw data rate (Mbps) | Spectrum Efficiency | Raw data rate (Mbps) | Spectrum Efficiency | Raw data rate (Mbps) | Spectrum Efficiency |
| QPSK | 1/2 | 15.5844 | 0.7792 | 15.1261 | 0.7563 | 14.2857 | 0.7143 | 12.8571 | 0.6429 |
| QPSK | 3/4 | 23.3766 | 1.1688 | 22.6891 | 1.1345 | 21.4286 | 1.0714 | 19.2857 | 0.9643 |
| 16QAM | 1/2 | 31.1688 | 1.5584 | 30.2521 | 1.5126 | 28.5714 | 1.4286 | 25.7143 | 1.2857 |
| 16QAM | 3/4 | 46.7532 | 2.3377 | 45.3782 | 2.2689 | 42.8571 | 2.1429 | 38.5714 | 1.9286 |
| 64QAM | 1/2 | 46.7532 | 2.3377 | 45.3782 | 2.2689 | 42.8571 | 2.1429 | 38.5714 | 1.9286 |
| 64QAM | 2/3 | 62.3377 | 3.1169 | 60.5942 | 3.0252 | 57.1429 | 2.8571 | 51.4286 | 2.5714 |
| 64QAM | 3/4 | 70.1299 | 3.5065 | 68.0672 | 3.4034 | 64.2857 | 3.2143 | 57.8571 | 2.8929 |

## GLOSSARY

| | |
|---|---|
| B | Total amount of bandwidth available at the BS for uplink Connections (Mbps) |
| W | Bandwidth in Hz |
| $B_U$ | Minimum Reserved Rate for UGS connections (kbps) |
| $B_r^{min}$ | Minimum Reserved Rate for rtPS connections (kbps) |
| $B_r^{max}$ | Maximum Sustained Rate for rtPS connections (kbps) |
| L | Maximum Latency for rtPS connections (ms) |
| $B_n^{min}$ | Minimum Reserved Rate for nrtPS connections (kbps) |
| $B_n^{max}$ | Maximum Sustained Rate for nrtPS connections (kbps) |
| $MRTR_i$ | Minimum Reserved Traffic Rate of connection type i |
| f | Duration of a timeframe which includes downlink and uplink subframes (ms) |
| $r_i$ | Token arrival rate of a connection type i (kbps) |
| $b_i$ | Token bucket size of a connection type i (kbits) |
| $m_i$ | L/f , $m_i$ must be an integer |
| $E_b/N_{0,i}$ | Signal energy per bit to noise density of connection type i |
| $SINR_i$ | Measured signal to interference plus noise ratio of a connection type i |
| $SINR_{th,i}$ | Calculated threshold signal to interference plus noise ratio of a connection type i |
| $B_{rem}$ | Amount of bandwidth left after bandwidth allocation to admitted connection |
| $B_{req,i}$ | Bandwidth request of a connection type i |
| $C_{NRT}$ | Total amount of bandwidth allocated to non real-time connections. |
| $C_{rtPS}$ | Total amount of bandwidth allocated to rtPS connections. |
| $n_u$ | Number of UGS connections admitted into the network |
| $n_r$ | Number of rtPS connections admitted into the network |
| $n_n$ | Number of nrtPS connections admitted into the network |
| $d_r$ | Current Degraded Bandwidth of rtPS connections |
| $d_n$ | Current Degraded Bandwidth of nrtPS connections |
| $B_{poll}$ | Remaining uplink bandwidth allotted for the polling services |
| $B_{total}^r$ | Remaining uplink bandwidth for real-time traffic flows |
| $B_{total}^n$ | Remaining uplink bandwidth for non real-time traffic flows |

### C. Description of the SINR based CAC algorithm

In the proposed CAC mechanism, the QoS requirements under consideration are indicated by SINR value, transmission delay and bandwidth availability of the connections. The proposed CAC algorithm uses the Signal to Interference plus Noise Ratio (SINR) to check for the availability of enough OFDMA sub-carriers for the new connection request. When a new connection request arrives, the BS calculates the SINR threshold (SINRth) for the requesting connection. The BS then updates the measured SINRi for each connection type, with the assumption that requesting connection of service type 'i' is in the system by occupying an OFDMA sub-carrier. If the measured SINRi is greater than or equal to the corresponding SINRth, and the transmission delay requirement is guaranteed for every real time connection in the system and also the required bandwidth is available for the requesting connection, the connection is admitted into the system. However, if the former two conditions are satisfied but bandwidth availability is not met, the requesting connection is admitted in the system through adaptive bandwidth degradation. Otherwise, the connection request is rejected. The detail of CAC mechanism is shown in the Figure 1.

### D. Calculation of SINRth and SINRi

The SINRth for the requesting connection should be determined in such a way so that the wireless transmission error remains below the acceptable threshold for the connection type 'i'. To find SINRth, the effective bit rate of the requesting connection type is first determined. The effective bit rate is defined as the minimum bandwidth required by traffic source to meet the transmission error threshold. In this paper, we have considered that the effective bit rate of a connection type 'i', is its MRTR, as defined in the WiMAX standard. Hence, SINR threshold of a connection type 'i' can be calculated as given in equation (6).

$$SINR_{th,i} = (E_b/N_0)_i * (MRTR_i/W) \qquad (6)$$

Let the requesting connection be the Cpth connection and the requesting connection's traffic type be 'i' and the connections which are already admitted in the particular WiMAX cell is denoted by 'Cj', where $1 \le j \le p-1$ and $j \ne p$. To find the effects of the requesting connection on all other admitted connections, the BS measures the total power 'ψ' received from all existing connections 'Cj' in the system. The received power from all existing connection acts as interference to the requesting connection. Let 'Sj' be the received power from the existing connection which acts as interference to the OFDMA sub-carrier of the requesting connection and 'Si' be the signal power of the ith type of the Cpth requesting connection in an OFDMA sub-carrier as shown in figure 2. Hence,

$$\psi = \sum_{j=1}^{p-1} S_j \qquad (7)$$

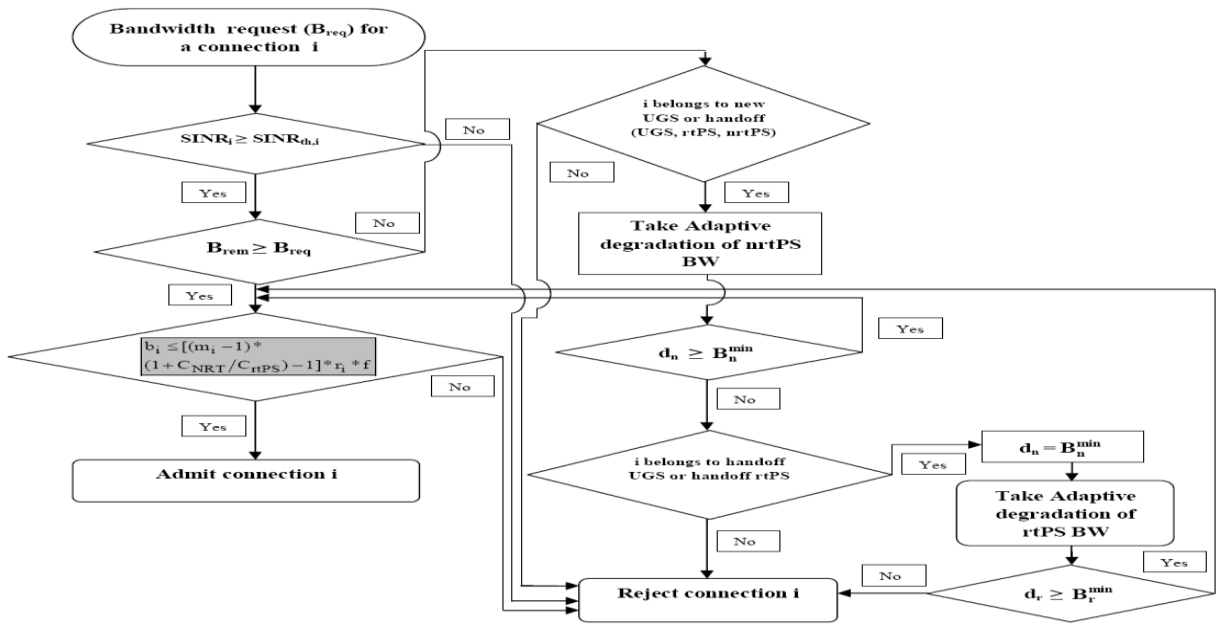

Figure 1.    Flow chart of proposed SINR based CAC algorithm for WiMAX BWA Networks

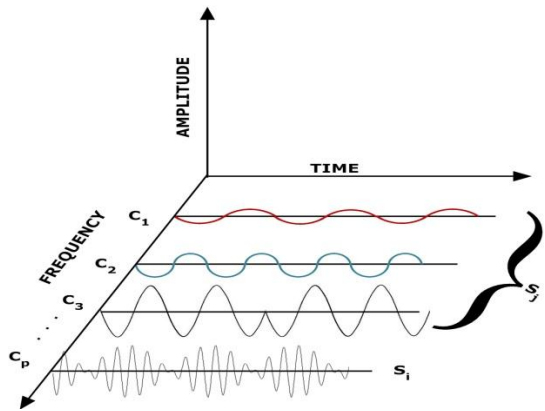

Figure 2.    OFDMA sub-carriers in time and frequency domain

Let $SINR_i$ be the new measured SINR for a connection when a new connection type 'i' wants to get admitted in the network and 'η' be the thermal noise in the network. $SINR_i$ is updated by the BS using equation (10) as derived below [10].

$$SINR_i = \frac{S_i}{\psi + \eta}$$

$$\Rightarrow SINR_i = \frac{S_i/\eta}{\frac{\psi}{\eta} + 1}$$

$$\Rightarrow SINR_i = \frac{S_i/\eta}{\frac{\sum_{j=1}^{p-1} S_j}{\eta} + 1} \qquad (8)$$

Again $S_i/\eta$ for a connection can be calculated as given in [37],

$$\frac{S_i}{\eta} = \left(\frac{E_b}{N_0}\right)_i * \frac{r_i}{W} \qquad (9)$$

Replacing $\frac{S_i}{\eta} = \left(\frac{E_b}{N_0}\right)_i * \frac{r_i}{W}$ in (8) we get,

$$SINR_i = \frac{\left(\frac{E_b}{N_0}\right)_i * \left(\frac{r_i}{W}\right)}{\frac{\sum_{j=1}^{p-1} S_j}{\eta} + 1} \qquad (10)$$

Where '$r_i$' is the transmission data rate of $i^{th}$ connection type.

Again $\frac{\sum_{j=1}^{p-1} S_j}{\eta}$ can be calculated using equation (9) for all connection type,

$$\frac{\sum_{j=1}^{p-1} S_j}{\eta} = (n_u * \left(\frac{E_b}{N_0}\right)_U * \frac{r_U}{W}) + (n_r * \left(\frac{E_b}{N_0}\right)_r * \frac{r_r}{W}) + (n_n * \left(\frac{E_b}{N_0}\right)_n * \frac{r_n}{W}) \quad (11)$$

Where '$r_u$', '$r_r$' and '$r_n$' are the bit rates and '$n_u$', '$n_r$', '$n_n$' are the number of admitted connection of UGS, rtPS and nrtPS services respectively.

### E.  Description of delay guarantee condition

A condition has been provided in [18] to satisfy the Delay Guarantee required by the rtPS connection type 'i'. Since token bucket mechanism is used to schedule the packets, maximum number of packets (in terms of arriving bits) that arrive in a time frame of duration f is $b_i + r_i * f$. These arriving bits must be scheduled in next ($m_i$-1) time frame (see Table III) to avoid delay violation. The required condition is given below, (for proof of the condition kindly see [18])

$$b_i \leq [(m_i - 1) * (1 + C_{NRT,i}/C_{rtPS,i}) - 1] * r_i * f \qquad (12)$$

Where $C_{NRT,i} = B - (n_U * B_U) - (n_r * B_r)$ and $C_{rtPS,i} = n_r * B_r$ .

Here $C_{NRT,i}$ and $C_{rtPS,i}$ are calculated taking the new connection type '*i*' to be admitted. Here, '*i*' refers to the rtPS connection type.

### F. Adaptive bandwidth degradation of rtPS and nrtPS connections

In this scheme, bandwidths of lower priority connections are degraded as per minimum bandwidth requirement of the newly admitted or handoff connections. To admit more new UGS connections, the bandwidth is degraded only from nrtPS connections. In addition, degradation is performed on both rtPS and nrtPS connections to allow more UGS, rtPS and nrtPS handoff calls. The advantages of adaptive bandwidth degradation [38] over fixed bandwidth degradation [12- 16] are:

- No fixed step size degradation.
- No need to assign initial arbitrary step size.
- Instead, the minimum required bandwidth is calculated and then degraded.
- The degradation is adaptive to the required bandwidth.
- Bandwidth utilization of the system is greatly improved.

### G. Need for rescheduling of bandwidth at SS

According to IEEE 802.16 standard, BS is responsible for allocating the uplink bandwidth based on the request from SSs. As the SS may have multiple connections, the bandwidth request message should report the bandwidth requirement of each connection to BS. All packets from application layer in the SS are classified by the connection classifier based on CID and are forwarded to the appropriate queue. Scheduler at SS will retrieve the packets from the queues and transmit them to the network in the appropriate time slots, as defined by the uplink map massage (UL-MAP) [18] sent by the BS. The UL-MAP is determined by the uplink packet scheduling module based on the BW-request messages that report queue size of each connection in SS. But the scheduler inside the BS may have only limited or even outdated information about the current state of each uplink connection due to large Round Trip Delay (RTD) [25] as shown in Figure 3.

So a scheduling algorithm is needed in each SS to reassign the received transmission bandwidth among different connections. Since the uplink traffic is generated at SS, the SS scheduler is able to arrange the transmission bandwidth on the up-to-date information and provide tight QoS guarantee for its connections.

Hereunder, we integrate the distributed scheduling algorithm [26], with our proposed SINR based CAC scheme. The scheduling algorithm is based on current queue size and queue delay at SS.

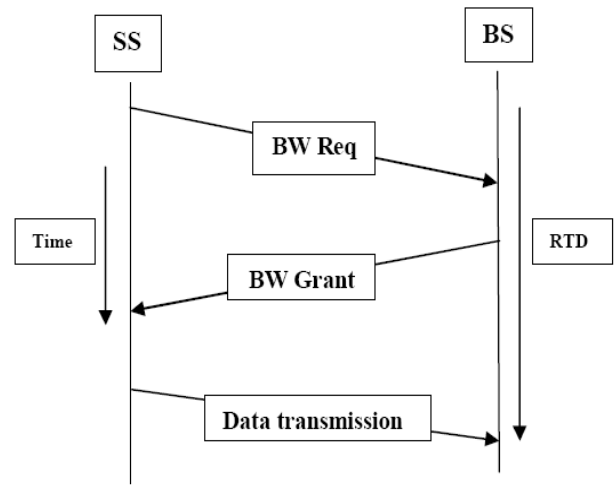

Figure 3.   RTD for Uplink transmission in IEEE 802.16

### H. Scheduling algorithm at SS

Since UGS service has a critical delay and delay jitter requirement and its transmission cannot be deferred or interrupted by other flows, SS scheduler firstly guarantees the bandwidth for UGS queue. This generally takes a fixed chunk of bandwidth. The remaining uplink bandwidth 'Bpoll' that is allotted for the polling service will be

$$B_{poll} = B - n_U * B_U \qquad (13)$$

A parameter α, proposed in [26], is defined as the ratio of the maximum time a rtPS or nrtPS MAC Protocol Data Unit (MPDU) can wait in the queue (i.e. max_mpdu_delay) to the maximum latency specification of the real-time flows.

$$\alpha = \frac{\text{max\_ mpdu\_ delay}}{\text{max\_ latency\_ of \_rtPS\_ flow}} \qquad (14)$$

This 'α' can be considered as a design parameter which provides the present status of the queue. It also controls the QoS of the real and non-real time services. The remaining uplink bandwidth after allocation to UGS, are divided among 'nr' and 'nn' numbers of real-time and non real-time flows as follows [1, 26].

The real-time traffic flows are allotted an uplink bandwidth of

$$B_{tot}^r = B_{poll} * \frac{n_r * \alpha}{n_r * \alpha + n_n} \qquad (15)$$

The non real-time traffic flows are allotted an uplink bandwidth of

$$B_{tot}^n = B_{poll} * \frac{n_r}{n_r * \alpha + n_n} = B_{poll} - B_{tot}^r \qquad (16)$$

Every SS repeats this process at the beginning of every uplink. Since the bandwidth request/grant will take some time due to RTD, the rtPS traffic can actually vary within this period. This algorithm is not affected by the rtPS delay bound traffic because the granted bandwidth is always redistributed.

In this way this algorithm provides fair distribution of the bandwidth.

*I. Cross layer Architecture*

Figure 4 shows the cross layer architecture proposed in this paper. At PHY layer FFT Size =2048 and Cyclic Prefix=1/32 have been selected because these parameters provide highest raw data rate and spectrum efficiency for all types of MCS as shown in Table II. In addition, various MCS like QPSK-1/2, QPSK-3/4, 16QAM-1/2, 16QAM-3/4, 64QAM-1/2, 64QAM-2/3 and 64QAM-3/4 have been selected at PHY layer to evaluate their individual performance in IEEE 802.16 networks.

Further, our proposed CAC algorithm and the Scheduling algorithm have been implemented at MAC layer of BS and SS respectively. The input parameters for the CAC algorithm are the SINR for all the connections, a condition for providing delay guarantee to the rtPS connection and availability of bandwidth. Based on these input parameters, CAC algorithm takes the proper decision to admit or reject an incoming connection request along with its required bandwidth. The output performance parameters obtained from the CAC algorithm are NCBP, HCDP and COP. These output performance parameters act as feedback information, in order to select appropriate MCS dynamically at PHY layer to support the required QoS guarantee.

This dynamic allocation of MCS at PHY layer based on feedback information from MAC layer employs Adaptive Modulation and Coding (AMC) technique in the network. Scheduling algorithm takes current queue size and queuing delay requirement as the input parameters to reschedule the bandwidth shared between real-time and non real-time traffic. It can be noticed that bandwidth which has already been granted or scheduled by the CAC algorithm at BS is again rescheduled by the Scheduling algorithm at SS among the traffic sources. Bandwidth utilization and fairness index are the output performance parameters obtained from the Scheduling algorithm which reveals efficient utilization and fair allocation of the system resources for all kind of MCS. Thereby, when AMC technique selects a MCS in the network, the scheduling algorithm distribute the bandwidth provided by that MCS among the various traffic sources in the network. In this way, Scheduling algorithm helps to improve utilization and fair allocation of the system resources for all MCS.

### III. ANALYTICAL MODEL

In this paper, the performance evaluation of the CAC mechanisms is obtained by using the Continuous Time Markov Chain (CTMC) Model [35]. The Markov model has been opted because it examines the probability of being in a given state at a given point of time, the amount of time a system is expected to spend in a given state, as well as the expected number of transitions between states.

A single WiMAX BS in isolation is considered. This BS will receive the bandwidth requests from the SS within the coverage area of a BS. Three types of services UGS, rtPS, nrtPS need QoS guarantees and request for connection admission. BS changes state from one to another upon the admission or rejection of a connection. Further, it is assumed that the BS either admits or rejects only one connection at a particular instant of time. So the next state of the BS depends only on the present state of the BS but does not depend on the previous states of the BS.

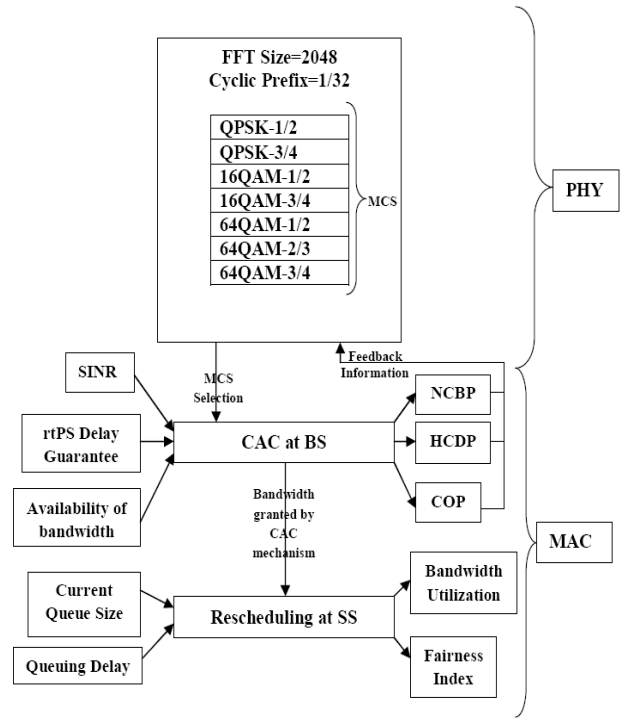

Figure 4. Cross layer architecture for IEEE 802.16 BWA network

Therefore, the states of the BS form a Markov Chain, and accordingly the BS can analytically be modeled as shown in Figure 5. In this scenario, the BS can uniquely be represented in the form of a five dimensional Markov Chain $(n_u, n_r, d_r, n_n, d_n)$ based on the number of admitted connections of each type.

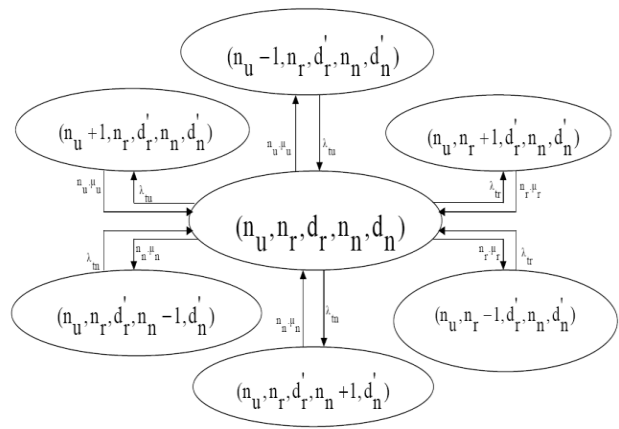

Figure 5. State Transition Diagram of the Markov Chain model for proposed CAC Algorithm

State $s = (n_u, n_r, d_r, n_n, d_n)$ represents that the BS has currently admitted '$n_u$', '$n_r$', and '$n_n$' number of UGS, rtPS and nrtPS connections respectively into the network. In Figure 5, '$d_r^{'}$' and '$d_n^{'}$' are the degraded bandwidth of rtPS and nrtPS connections respectively after state transition and may have a different value from that of '$d_r$' and '$d_n$'. The BS will be in a particular State $s = (n_u, n_r, d_r, n_n, d_n)$ until a new connection of one of them i.e. UGS, rtPS, nrtPS is admitted into the network or an ongoing connection is terminated. The arrival process of the handoff and newly originated UGS, rtPS, and nrtPS connections is Poisson with rates $\lambda_{hu}$, $\lambda_{hr}$, $\lambda_{hn}$, $\lambda_{ou}$, $\lambda_{or}$, and $\lambda_{on}$ respectively. Therefore Total arrival rate $\lambda_{tu}$, $\lambda_{tr}$, $\lambda_{tn}$ corresponding to UGS, rtPS and nrtPS connections have been evaluated as given below,

$$\lambda_{tu} = \begin{cases} \lambda_{hu} + \lambda_{ou}, \text{ if}(n_u+1)B_U + n_r B_r^{max} + n_n B_n^{min} \le B \\ \\ \lambda_{hu}, \quad \text{otherwise} \end{cases} \quad (17)$$

$$\lambda_{tr} = \begin{cases} \lambda_{hr} + \lambda_{or}, \text{ if } n_u B_U + (n_r+1)B_r^{max} + n_n B_n^{min} \le B \\ \\ \lambda_{hr}, \quad \text{otherwise} \end{cases} \quad (18)$$

$$\lambda_{tn} = \begin{cases} \lambda_{hn} + \lambda_{on}, \text{ if } n_u B_U + n_r B_r^{max} + (n_n+1)B_n^{max} \le B \\ \\ \lambda_{hn}, \quad \text{otherwise} \end{cases} \quad (19)$$

The service times of UGS, rtPS and nrtPS connections are exponentially distributed with mean $1/\mu_u$, $1/\mu_r$ and $1/\mu_n$ respectively.

The state space **S** for our proposed CAC scheme is obtained based on the following equation.

$$\mathbf{S} = \{ \qquad s = (n_u, n_r, d_r, n_n, d_n) \mid$$

$$(n_u.B_U + n_r.d_r + n_n.d_n) \le B$$
$$\wedge (B_r^{min} \le d_r \le B_r^{max}) \wedge (B_n^{min} \le d_n \le B_n^{max}) \} \quad (20)$$

From Figure 5, it is observed that every state $s = (n_u, n_r, d_r, n_n, d_n)$ in state space '**S**' is reachable from every other state i.e. each state communicates with other states in the state space '**S**'. As for example, state $(n_u, n_r, d_r, n_n, d_n)$ communicates with state $(n_u, n_r+1, d_r^{'}, n_n, d_n^{'})$ and also communicates with state $(n_u, n_r, d_r^{'}, n_n-1, d_n^{'})$. Hence state

$(n_u, n_r+1, d_r^{'}, n_n, d_n^{'})$ and $(n_u, n_r, d_r^{'}, n_n-1, d_n^{'})$ can also communicate to each other. In this way, each state can communicate with other state in the state space '**S**'. Therefore, the state space '**S**' forms a closed set and the Markov chain obtained is irreducible [35].

Let the steady state probability of the state $s = (n_u, n_r, d_r, n_n, d_n)$ is represented by $\pi_{(n_u, n_r, d_r, n_n, d_n)}$. As the Markov chain is irreducible, thereby observing the outgoing and incoming states for a given state '$s$', the state balance equation of state s is shown in equation (21).

$$(\lambda_{tu}\varphi_{(v+1,w,x,y,z)} + \lambda_{tr}\varphi_{(v,w+1,x,y,z)} + \lambda_{tn}\varphi_{(v,w,x,y+1,z)} +$$
$$v\mu_u\varphi_{(v-1,w,x,y,z)} + w\mu_r\varphi_{(v,w-1,x,y,z)} + y\mu_n\varphi_{(v,w,x,y-1,z)})\pi_{(v,w,x,y,z)} =$$
$$\lambda_{tu}\varphi_{(v-1,w,x',y,z')}\pi_{(v-1,w,x',y,z')} + \lambda_{tr}\varphi_{(v,w-1,x',y,z')}\pi_{(v,w-1,x',y,z')}$$
$$+ \lambda_{tn}\varphi_{(v,w,x',y-1,z')}\pi_{(v,w,x',y-1,z')} + (v+1)\mu_u\varphi_{(v+1,w,x',y,z')}\pi_{(v+1,w,x',y,z')}$$
$$+ (w+1)\mu_r\varphi_{(v,w+1,x',y,z')}\pi_{(v,w+1,x',y,z')}$$
$$+ (y+1)\mu_n\varphi_{(v,w,x',y+1,z')}\pi_{(v+1,w,x',y+1,z')} \quad (21)$$

Where v, w, x, y and z represent $n_u$, $n_r$, $d_r$, $n_n$ and $d_n$ respectively.

$\varphi_{(v,w,x,y,z)}$ represents the characteristic equation as shown below.

$$\varphi_{(v,w,x,y,z)} = \begin{cases} 1, (v,w,x,y,z) \in \mathbf{S} \\ \\ \\ 0, \quad \text{otherwise} \end{cases}$$

By using equation (21), the state balance equations of each state in the state space '**S**' are obtained. Solutions of these equations provide the steady state probabilities of all states in the space S with the normalized condition imposed by equation (22).

$$\sum_{s \in S} \pi_{(n_U, n_r, d_r, n_n, d_n)}(s) = 1 \quad (22)$$

From the steady state probabilities we can determine various QoS parameters of the system as given under.

### A. New Connection Blocking Probability (NCBP)

The new connection blocking probability is the probability of rejecting a new connection request for admission into the network. Conditions for blocking a new connection request have been included in Table IV.

- *Estimation of NCBP for UGS, rtPS, nrtPS connections*

While admitting a new connection, if the next state is not allowable Markov chain state in the state space '**S**' due to the conditions as given in Table IV, then the next state is considered to be a blocked state for that connection i.e. the new connection is blocked.

Let '$S_{UB}$', '$S_{rB}$' and '$S_{nB}$' form a state space for the states whose next state are not allowed in the Markov chain due to connection blocking.

The summation of the steady state probabilities of the states in the state space '$S_{UB}$', '$S_{rB}$' and '$S_{nB}$' give the NCBP of UGS, rtPS and nrtPS connections respectively. Hence,

$$NCBP\text{-}UGS = \sum_{s \in S_{UB}} \pi_{(n_u, n_r, d_r, n_n, d_n)}(s) \qquad (23)$$

$$NCBP\text{-}rtPS = \sum_{s \in S_{rB}} \pi_{(n_u, n_r, d_r, n_n, d_n)}(s) \qquad (24)$$

$$NCBP\text{-}nrtPS = \sum_{s \in S_{nB}} \pi_{(n_u, n_r, d_r, n_n, d_n)}(s) \qquad (25)$$

*B. Handoff Connection Dropping Probability (HCDP):*

The handoff connection dropping probability is the probability of rejecting a handoff connection request for admission into the network. Conditions for dropping a handoff connection request are summarized in Table IV.

- *Estimation of HCDP for UGS, rtPS and nrtPS connections*

Again, for the handoff connection if the conditions as given in Table IV occur, then the handoff connection is dropped and the next state is not allowed in the Markov chain due to connection dropping.

Similar to NCBP, HCDP can also be estimated as below, where '$S_{UD}$', '$S_{rD}$' and '$S_{nD}$' form a state space corresponding to UGS, rtPS and nrtPS connection for the states whose next state is not allowed in the Markov chain due to connection dropping. Hence,

$$HCDP\text{-}UGS = \sum_{s \in S_{UD}} \pi_{(n_u, n_r, d_r, n_n, d_n)}(s) \qquad (26)$$

$$HCDP\text{-}rtPS = \sum_{s \in S_{rD}} \pi_{(n_u, n_r, d_r, n_n, d_n)}(s) \qquad (27)$$

$$HCDP\text{-}nrtPS = \sum_{s \in S_{nD}} \pi_{(n_u, n_r, d_r, n_n, d_n)}(s) \qquad (28)$$

*C. Connection Outage Probability (COP)*

The connection outage probability is the probability that the SINR of the connections in the network drops below a certain threshold when a new connection wants to get admitted in the network. Condition for connection outage in the network is summarized in Table IV.

- *Estimation of COP for UGS, rtPS and nrtPS connection*

A new or handoff connection is not admitted in the network if SINR of the connection falls below its threshold value as given in Table IV.

Similar to NCBP and HCDP, COP can be estimated as shown below.

$$COP\text{-}UGS = \sum_{s \in S_{UO}} \pi_{(n_u, n_r, d_r, n_n, d_n)}(s) \qquad (29)$$

$$COP\text{-}rtPS = \sum_{s \in S_{rO}} \pi_{(n_u, n_r, d_r, n_n, d_n)}(s) \qquad (30)$$

$$COP\text{-}nrtPS = \sum_{s \in S_{nO}} \pi_{(n_u, n_r, d_r, n_n, d_n)}(s) \qquad (31)$$

As $E_b/N_0$ ratio of the connections has the direct relationship (e.g. Equation (6), (10) and (11)) to the SINR and $SINR_{th}$, so from the proposed CAC mechanism and above defined parameters it is observed that the $E_b/N_0$ ratio provides significant influence on NCBP, HCDP and COP for all the connection types. The effect of $E_b/N_0$ ratio on NCBP, HCDP and COP will help to determine the optimum range of $E_b/N_0$ ratio for all the connections, in order to select an appropriate MCS to improve the channel condition. This dynamic selection of MCS based on $E_b/N_0$ ratio of the connections, employs Adaptive Modulation and Coding (AMC) technique in the network.

*D. Bandwidth Utilization (BU):*

The Bandwidth Utilization (BU) is defined [16] as the ratio of total used bandwidth to the available bandwidth of the system. BU of the system is encountered to estimate whether SS lying in a bad channel state wastes precious bandwidth. BU can be obtained as follows.

$$BU = \frac{\sum_{s \in S} (n_u.B_U + n_r.d_r + n_n.d_n)\pi_{(v,w,x,y,z)}}{B} \qquad (32)$$

Where, 'B' is the total bandwidth.

*Fairness Index*

Fairness refers to the equal allocation of network resources among the various users operating in both good and bad channel states. Fairness is quantified using Jain's Fairness Index (JFI) [27] as given below.

$$JFI = \frac{(\sum_{i=1}^{n} r_i)^2}{n * \sum_{i=1}^{n} r_i^2} \qquad (33)$$

Where, $r_i$ is the data rate of connection type i.

TABLE III.        CONDITION FOR CONNECTION BLOCKING, DROPPING AND OUTAGE IN THE NETWORK

| Current State | Next State | Condition | Status |
|---|---|---|---|
| $(n_u, n_r, d_r, n_n, d_n)$ | $(n_u + 1, n_r, d_r, n_n, d_n^{'})$ | $d_n^{'} < B_n^{min}$ or $(n_u + 1)*B_U + n_r*B_r^{max} + n_n*B_n^{min} > B$ | New UGS blocked |
| | | $d_n^{'} < B_n^{min}$ or $d_r^{'} < B_r^{min}$ or $(n_u + 1)*B_U + n_r*B_r^{min} + n_n*B_n^{min} > B$ | Handoff UGS dropped |
| | | $(E_b/N_0)_u/(n_u*(E_b/N_0)_u + n_r*(E_b/N_0)_r + n_n*(E_b/N_0)_n) < SINR_{th,u}$ | UGS outage |
| | $(n_u, n_r + 1, d_r, n_n, d_n^{'})$ | $d_n^{'} < B_n^{min}$ or $(n_u*B_U + (n_r + 1)*B_r^{max} + n_n*B_n^{min} > B$ | New rtPS blocked |
| | | $d_n^{'} < B_n^{min}$ or $d_r^{'} < B_r^{min}$ or $n_u*B_U + (n_r + 1)*B_r^{min} + n_n*B_n^{min} > B$ | Handoff rtPS dropped |
| | | $(E_b/N_0)_r/(n_u*(E_b/N_0)_u + n_r*(E_b/N_0)_r + n_n*(E_b/N_0)_n) < SINR_{th,r}$ | rtPS outage |
| | $(n_u, n_r, d_r, n_n + 1, d_n)$ | $n_u*B_U + n_r*B_r^{max} + (n_n + 1)*B_n^{max} > B$ | New nrtPS blocked |
| | | $d_n^{'} < B_n^{min}$ or $n_u*B_U + n_r*B_r^{max} + (n_n + 1)*B_n^{min} > B$ | Handoff nrtPS dropped |
| | | $(E_b/N_0)_n/(n_u*(E_b/N_0)_U + n_r*(E_b/N_0)_r + n_n*(E_b/N_0)_n) < SINR_{th,n}$ | nrtPS outage |

## IV.    PARAMETERS FOR PERFORMANCE EVALUATION

The frame duration (f) is taken as 1ms because it provides more bandwidth utilization of the system compared to higher length of frame duration as analyzed in our earlier work [19]. Total channel bandwidth (W) is taken as 20 MHz because this is the maximum bandwidth supported by the WiMAX network [36]. The FFT size is taken as 2048 with cyclic prefix 1/32 as this combination provides highest raw data rate and spectrum efficiency for all types of MCS as calculated in Table II.

As the IEEE 802.16 standards have not specified values for the QoS parameters, we have considered the system parameters for performance evaluation as given in Table V. The arrival rates of all the connections are assumed to be same i.e. $\lambda_{hU} = \lambda_{oU} = \lambda_{hr} = \lambda_{or} = \lambda_{hn} = \lambda_{on}$. The service time for UGS, rtPS and nrtPS connections is exponentially distributed with mean $\frac{1}{\mu_U}, \frac{1}{\mu_r}, \frac{1}{\mu_n}$ respectively and we assume $\mu_U = \mu_r = \mu_n = 0.2$. That is, the mean service time of the connections is taken as 5 sec. Service time of the admitted connections should be as minimal as possible [16] in order to provide access into the network by other users. Degradation of rtPS and nrtPS connections is adaptive to the required bandwidth.

TABLE IV.    QoS PARAMETERS

| Service | Max.Sustained traffic rate (kbps) | Min. reserved traffic rate (kbps) | Bucket Size (Bits) | Delay (ms) | Eb/N0 (dB) |
|---|---|---|---|---|---|
| *UGS* | 256 | 256 | 64 | - | 3.6 |
| *rtPS* | 1024 | 512 | 10240 | 21 | 6.3 |
| *nrtPS* | 1024 | 256 | 10240 | - | 8.1 |

## V.    NUMERICAL RESULTS

To evaluate the effectiveness and efficiency of the proposed cross layer architecture, the IEEE 802.16 PHY and MAC layer protocols are analyzed using MATLAB under version 7.3. Exhaustive simulations are carried out.

Firstly, the performance of the CAC algorithm is verified by using the CTMC model under various MCS. The effect of $E_b/N_0$ ratio on our proposed CAC mechanism is observed. The optimum range of $E_b/N_0$ ratio is determined to select an appropriate MCS in order to employ AMC in the network.

Next, the impact of queue based scheduling algorithm is examined on bandwidth utilization and Jain's fairness index for different connection arrival rates under various MCS. Justifications behind all the numerical results have been provided.

### A. Impact of CAC Mechanism on QoS informationParameters

Figure 6 shows the blocking probability of UGS, rtPS and nrtPS connections for different MCS. Total arrival rate of the connections have been taken from equation (17), (18) and (19). It is observed that 64QAM-3/4 provides least blocking probability and QPSK-1/2 provides highest blocking probability to the new connections to be admitted in the network as compared to the other MCS. This is because of the highest raw data rate (70.1299 Mbps) and spectrum efficiency (3.5065) obtained by the 64QAM-3/4 among all other MCS (Reference Table II). Again, it is observed that performances of 16QAM-3/4 and 64QAM-1/2 have been merged together because of their same raw data rate.

Figure 7 shows the dropping probability of UGS, rtPS and nrtPS connections for different MCS. Here, it is also observed that, 64QAM-3/4 provides least dropping probability and QPSK-1/2 provides highest dropping probability to the ongoing hand off connections as compared to the other MCS because of the same reasons as mentioned previously.

Next, connection outage probability of UGS, rtPS and nrtPS are examined under different MCS and results are shown in Figure 8. Outage probability is the probability that the SINR of the connections in the network drops below a certain threshold when a new connection wants to get admitted in the network and thereby required QoS guarantee is not met to the admitted connections. It is observed that 64QAM-3/4 provides highest outage probability and QPSK-1/2 provides least outage probability as compared to other MCS. Since QAM scheme has the higher data rate and spectrum efficiency than the other MCS, so it has the capability of accommodating more number of connections. Thereby the SINR value of each admitted connections in the network gets decreased and falls below the threshold level, which results in the higher outage probability of the connections in the system.

Therefore, QAM schemes provide good QoS in terms of blocking probability of the newly admitted connections and dropping probability of the hand off connections but fail to provide good QoS in terms of connection outage probability of the network. On the other hand, QPSK schemes provide good QoS in terms of connection outage probability but fail to provide good QoS in terms of blocking and dropping probability.

Hence, there is a need for Adaptive Modulation and Coding (AMC) technique whose performance is based on feedback information of the MAC layer QoS parameters like connection blocking probability, connection dropping probability and connection outage probability. By dynamically changing the allocation of modulation and coding rate based on Eb/N0 ratio as well as SINR of the connections, AMC enables a flexible use of the network resources that can support nomadic or mobile operations.

### B. Impact of Eb/N0 ratio on QoS information Parameters

To comprehend how modulation and coding schemes could be made adaptive in a region under WiMAX cell boundary, the proposed SINR based CAC scheme is again simulated under various MCS. We consider Eb/N0 ratio of all the connections because it has the influence on both SINR and SINRth of the admitted connections in the network. Eb/N0 ratio is varied between 1 dB to 20 dB because within this span we are able to find out the adaptive range of all MCS. Connection arrival rate is kept constant at 10 calls per second.

Figure 9 shows the blocking probability under various MCS with different Eb/N0 ratio. Table VI shows the suitable range of Eb/N0 ratio where different MCS show zero blocking probability for all the connections. As for example, 64QAM-3/4 provides zero blocking probability in the range 4 dB to 20 dB for UGS and rtPS connection but for nrtPS connection the range is 9dB to 20 dB. So the optimum range of Eb/N0 ratio in which 64QAM-3/4 provides zero blocking probability is 9dB to 20 dB for all connections types. Again, we have taken 20 dB as the maximum value of the Eb/N0 ratio in our simulation so it is regardless to mention the maximum value of the Eb/N0 ratio because the MCS can provide zero blocking probability beyond this range.

Next, Figure 10 shows the dropping probability under various MCS with different Eb/N0 ratio, whereas Table VI also shows the range of Eb/N0 ratio where different MCS show zero dropping probability for all the connections in our proposed SINR based CAC scheme.

Figure 11 shows the outage probability under various MCS with different Eb/N0 ratio and also Table VI shows suitable range of Eb/N0 ratio in which different MCS show zero outage probability for all the connections.

TABLE V.  RANGE OF Eb/N0 RATIO

| Modulation and Coding Scheme | Minimum Eb/N0 ratio(dB) for zero blocking probability obtained from figure 9 | Minimum Eb/N0 ratio(dB) for zero dropping probability obtained from figure 10 | Maximum Eb/N0 ratio(dB) for zero outage probability obtained from figure 11 |
|---|---|---|---|
| *64QAM-3/4* | 9 | 4 | 1 |
| *64QAM-2/3* | 10 | 4 | 1 |
| *64QAM-1/2* | 14 | 6 | 2 |
| *16QAM-3/4* | 14 | 6 | 2 |
| *16QAM-1/2* | 20 | 9 | 3 |
| *QPSK-3/4* | - | 14 | 4 |
| *QPSK-1/2* | - | - | 7 |

By analyzing Table VI, it can be observed that, QAM schemes perform well at higher $E_b/N_0$ ratio in terms of zero blocking probability (range is above 9 dB ) and zero dropping probability (range is above 4 dB) and QPSK schemes perform well at lower $E_b/N_0$ ratio (range is below 7 dB) in terms of zero outage probability. However, zero probability is not achievable in the practical scenario owing to high intense traffic arrival rate in comparison to 10 calls per second which we have taken for our simulation. Despite of it, the evaluated adaptive range of $E_b/N_0$ ratio will provide minimum blocking, dropping and outage probability in the practical scenario.

Hence, it is better to select higher order MCS like QAM near to the WiMAX base station where SINR of the connections as well as population density of the users are more and lower order MCS like QPSK far away from the base station where SINR of the connections as well as population density of the users are less. This adaptive allocation of MCS makes the user to experience less blocking, less dropping and less outage in their connection request. Thus, AMC is employed in the network that enables dynamic use of the network resources to support nomadic or mobile operation. Therefore, cross layer adaptation of the different modulation capability at the PHY layer with the QoS requirement at the MAC layer is obtained.

*Impact of CAC Mechanism on QoS Information Parameters*

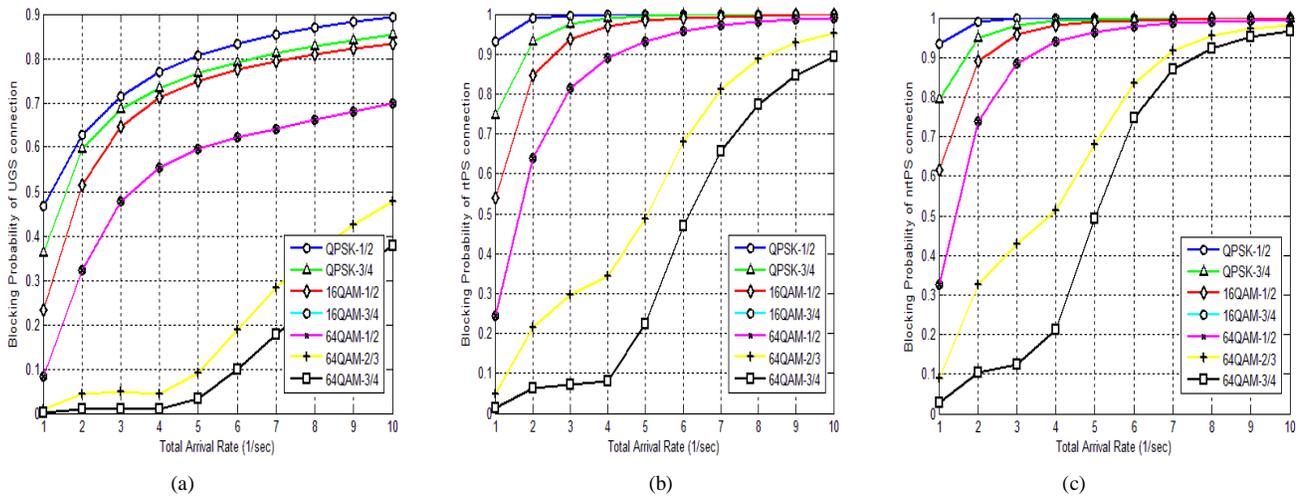

Figure 1. Blocking probability of UGS connections, (b) Blocking probability of rtPS connections and (c) Blocking probability of nrtPS connections

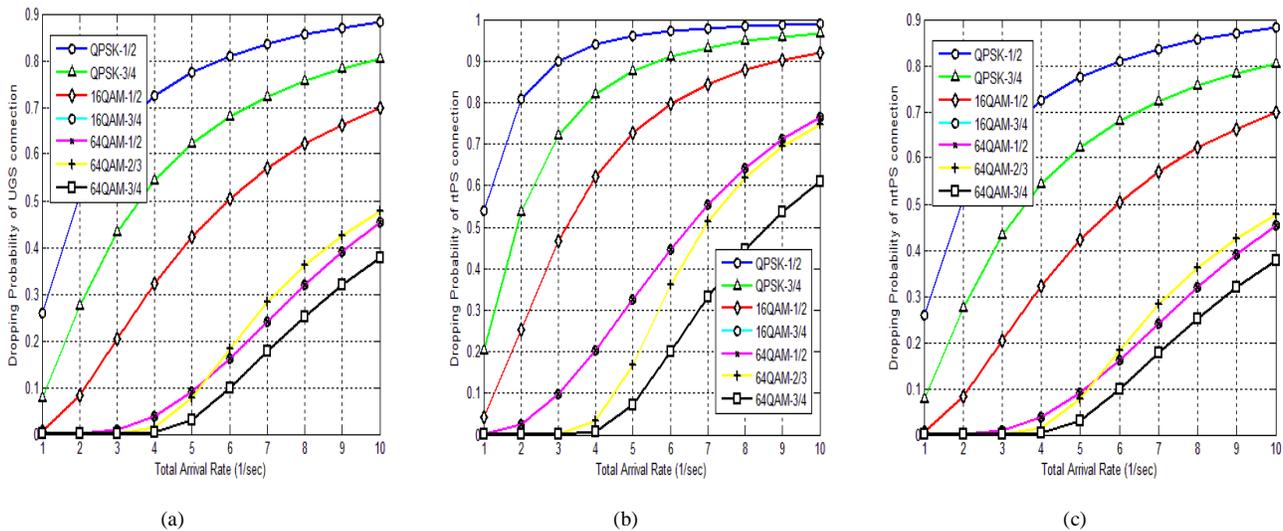

Figure 7(a) Dropping probability of UGS connections, (b) Dropping probability of rtPS connections and (c) Dropping probability of nrtPS connections

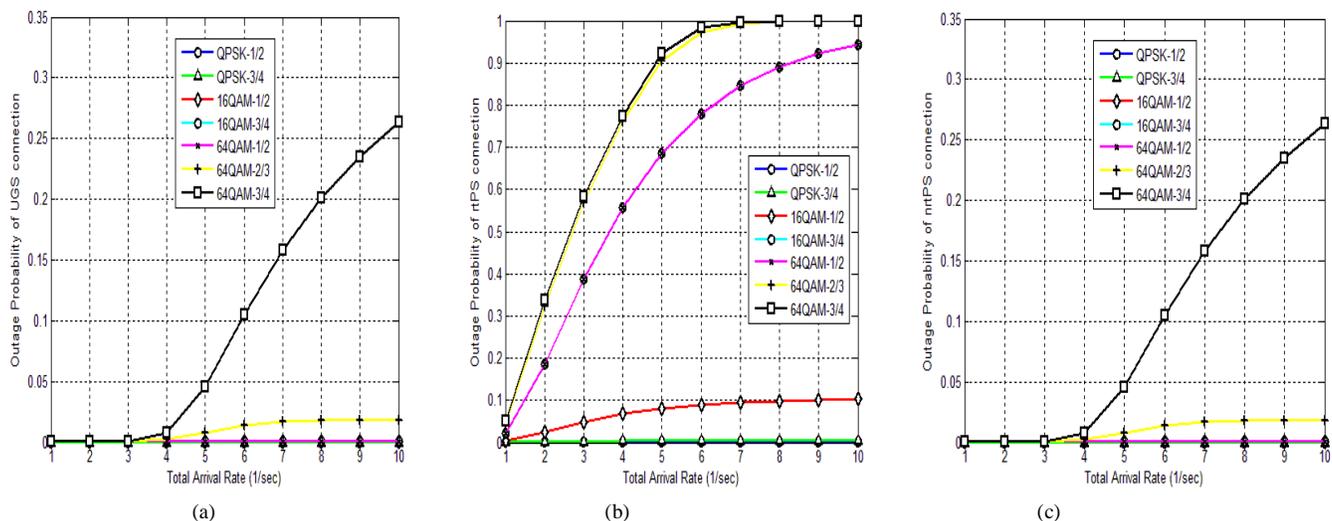

Figure 8(a) Outage probability of UGS connection, (b) Outage probability of rtPS connections and (c) Outage probability of nrtPS connection

***Impact of $E_b/N_0$ ratio on QoS Information Parameters***

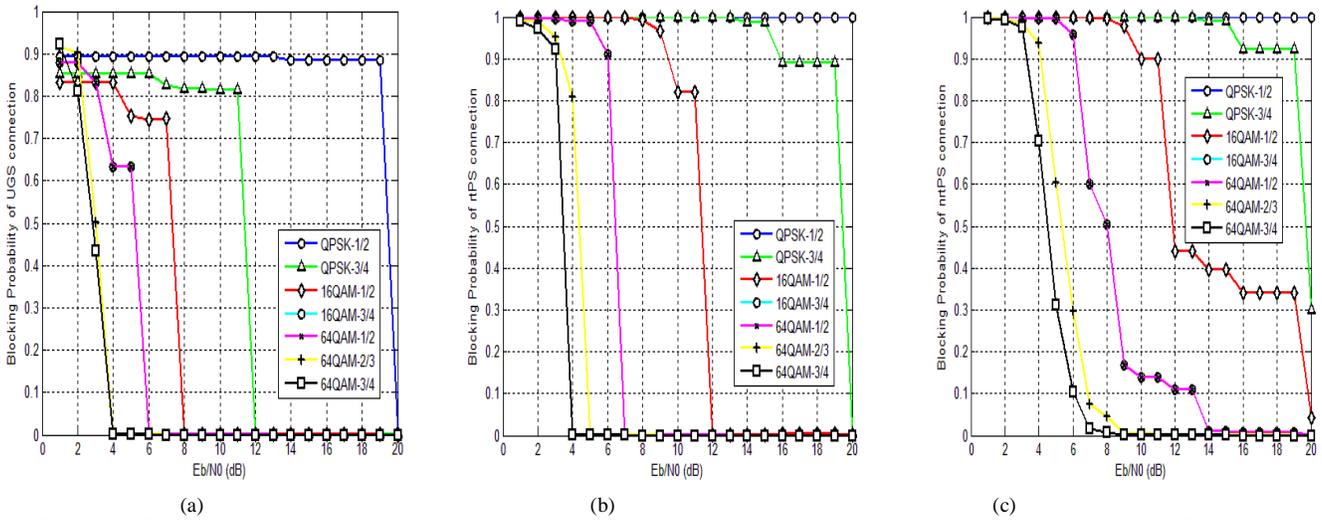

Figure 9 (a) Blocking probability of UGS connections, (b) Blocking probability of rtPS connections and (c) Blocking probability of nrtPS connection

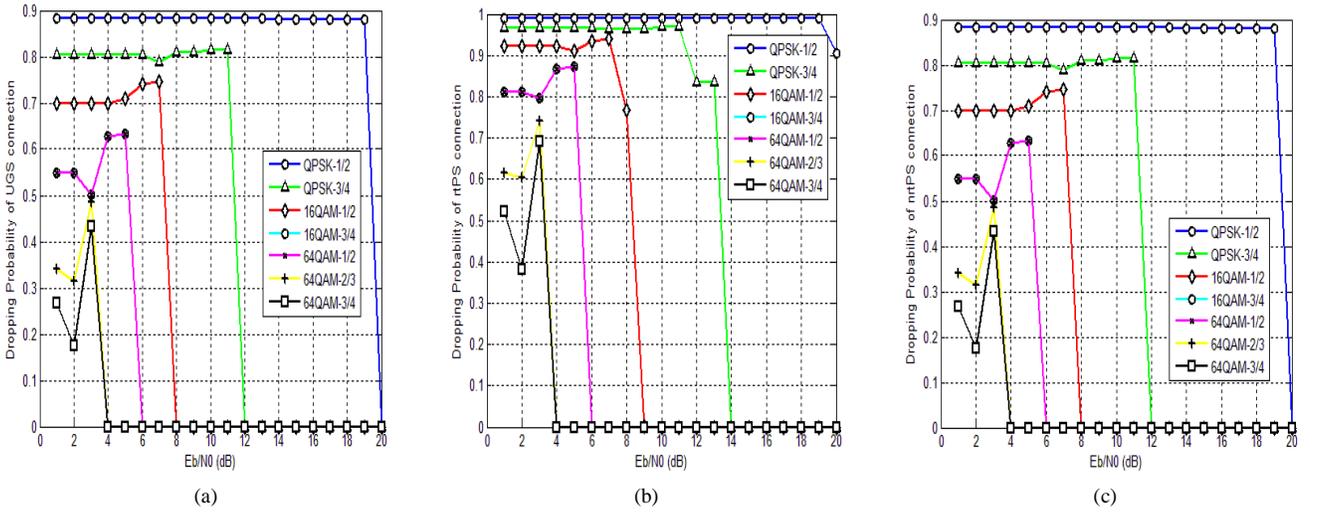

Figure 10(a) Dropping probability of UGS connections, (b) Dropping probability of rtPS connections and (c) Dropping probability of nrtPS connections

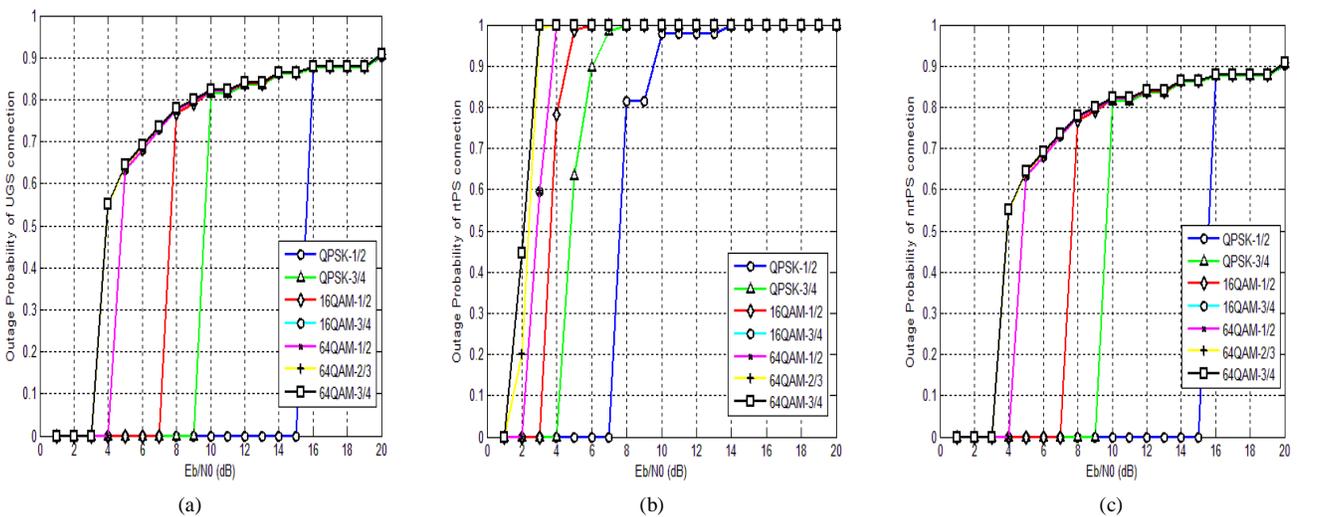

Figure 11(a) Outage probability of UGS connections, (b) Outage probability of rtPS connections and (c) Outage probability of nrtPS connections

## C. Impact of Queue Based Scheduling on QoS Parameters

To demonstrate the advantage of rescheduling of bandwidth in our proposed cross layer architecture, further simulation has been carried out for different connection arrival rates under various MCS.

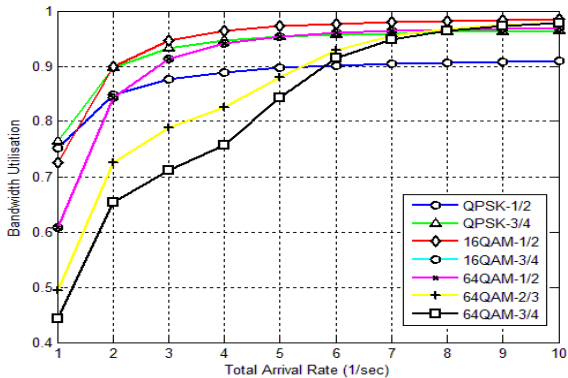

Figure 12(a). Average Bandwidth utilization before queue based rescheduling

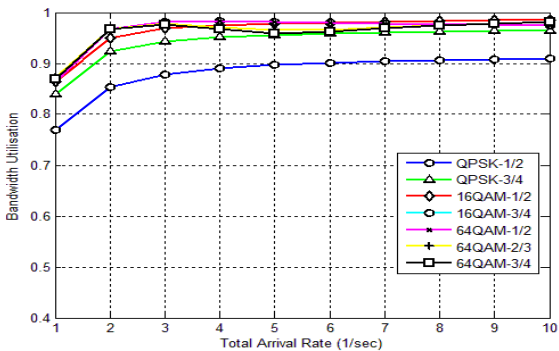

Figure 12 (b). Average Bandwidth utilization after queue based rescheduling

Figure 12(a) and 12(b) show the bandwidth utilization before and after queue based scheduling scheme respectively under different MCS. It shows the significant performance improvement in the bandwidth utilization due to rescheduling of the resource sharing between real-time and non real-time traffic based on their current queue size and latency requirements. From Figure 12(b), it can be observed that bandwidth utilization seems to be independent of MCS i.e. High bandwidth utilization for all MCS. It is because of the influence of scheduling algorithm at SS. As bandwidth is considered to be a limited resource in the network, rescheduling of bandwidth at SS will automatically improve the revenues of the service providers.

On the other hand, Figure 13(a) and 13(b) show the Jain's fairness index before and after the queue based scheduling scheme respectively under different MCS. It is also observed that substantial performance improvement in the fair allocation of resources between real-time and non real-time traffic is achieved because of the rescheduling of the resources. Hence, joint performance of the CAC and rescheduling algorithms also provides better resource utilization at lower traffic arrival rate.

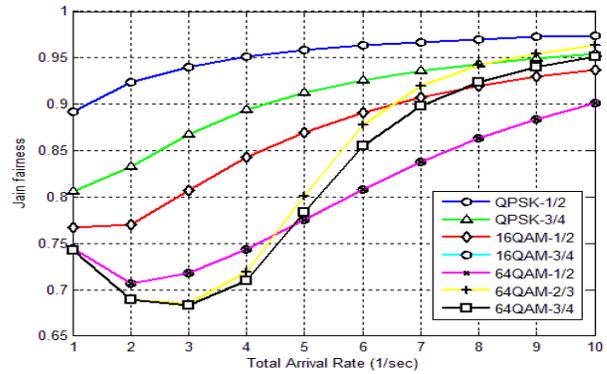

Figure 13(a). Jain's fairness index before queue based rescheduling

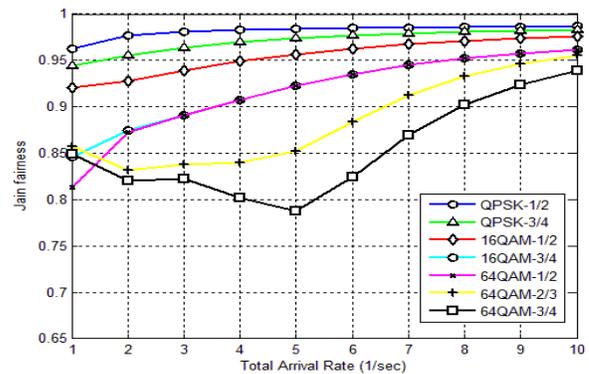

Figure 13(b). Jain's fairness index after queue based rescheduling

Percentage improvements of the performances after rescheduling of bandwidth for all MCS at traffic arrival rate = 1 are shown in Table VII. Highest performance improvement is observed in case of 64QAM-3/4 as compared to the other MCS. This is because 64QAM-3/4 has highest data rate and spectrum efficiency as calculated in Table II.

TABLE VI.     PERCENTAGE IMPROVEMENT IN QoS AFTER RESCHEDULING OF BANDWIDTH WHEN TOTAL ARRIVAL RATE = 1

| Modulation and Coding Scheme | Percentage improvement in the BU | Percentage improvement in the JFI |
|---|---|---|
| *QPSK-1/2* | 2.3813 | 7.4288 |
| *QPSK-3/4* | 10.0498 | 20.1068 |
| *16QAM-1/2* | 18.8653 | 29.4374 |
| *16QAM-3/4* | 43.2356 | 43.3220 |
| *64QAM-1/2* | 42.2646 | 45.8475 |
| *64QAM-2/3* | 76.5587 | 46.6461 |
| *64QAM-3/4* | 96.6757 | 49.1656 |

When AMC technique selects an MCS in the network based on the feedback information obtained from the CAC algorithm, the scheduling algorithm redistribute the bandwidth of that selected MCS among the various traffic sources in the network. In this way, scheduling algorithm helps to improve utilization and fair allocation of the system resources for all MCS in IEEE 802.16 BWA networks.

## VI. CONCLUSION

Cross layer adaptations are essential for guaranteeing QoS supports in real-time multimedia traffic over wireless networks. In this paper, a cross layer architecture for adapting different MCS in PHY layer has been considered by incorporating MAC layer information in terms of New Connection Blocking Probability (NCBP), Hand off Connection Dropping Probability (HCDP) and Connection Outage Probability (COP) for WiMAX BWA systems. In literatures, many researchers have proposed various cross layer mechanisms for providing better QoS support to the system, but so far no such comprehensive cross layer design considering the parameters stated above, has yet been reported in the literature. In this work, SINR based CAC integrated with the Queue based Scheduling has been analysed with Markov Chain model. The effect of $E_b/N_0$ ratio is observed on NCBP, HCDP and COP for all the connection types under various MCS by means of exhaustive simulation. Also optimum range of $E_b/N_0$ ratio is determined, in order to select an appropriate MCS which may be used as the threshold parameters for SINR in any adaptive modulation scheme. Moreover, the joint performance of the CAC and Scheduling algorithms has been proved to be good enough to meet the QoS requirements in terms of bandwidth utilization and Jain's fairness index.

### ACKNOWLEDGEMENT

The authors deeply acknowledge the support from DST, Govt. of India for this work in the form of FIST 2007 Project on "Broadband Wireless Communications" in the Department of ETCE, Jadavpur University.

AUTHORS PROFILE

1.  *Mr. Prasun Chowdhury* (prasun.jucal@gmail .com) has completed his Masters in Electronics and Telecommunication Engineering from Jadavpur University, Kolkata, India in 2009. Presently he is working as Senior Research Fellow (SRF) in the Department of Electronics and Telecommunication Engineering, Jadavpur University, Kolkata, India. His current research interests are in the areas of Call Admission control and packet scheduling in IEEE 802.16 BWA Networks. He has authored some journals and international conference papers.

2.  *Dr. Iti Saha Misra* (itimisra@cal.vsnl.net.in) received her PhD in Microstrip Antennas from Jadavpur University (1997). She is presently a Professor in the Department of Electronics and Telecommunication Engineering, Jadavpur University, India. Her current research interests include Mobility and Location Management, Next Generation Wireless Network Architecture and protocols, Call Admission control and packet scheduling in cellular and WiMAX networks. She has authored more than 100 research papers in refereed Journal and International Conference and a book on Wireless Communications and Networks. She is an IEEE Senior Member and founder Chair of the Women In Engineering, Affinity Group, IEEE Calcutta Section.

3.  *Dr. Salil K. Sanyal* (s_sanyal@ieee.org) received his Ph.D from Jadavpur University, India (1990). He is currently a Professor in the Department of Electronics and Telecommunication Engineering, Jadavpur University. He has authored more than 130 Research Papers in refereed Journals and International/National Conference Proceedings and also co-authored the Chapter "Architecture of Embedded Tunable Circular Microstrip Antenna" in the book entitled "Large Scale Computations, Embedded Systems and Computer Security". He is a Senior Member of IEEE and past Chair of IEEE Calcutta Section. His current research interests include Analog/Digital Signal Processing, VLSI Circuit Design, Wireless Communication and Tunable Microstrip Antenna.